\documentclass{emulateapj}
\usepackage{graphicx}
\begin{document}

\def\ba{\begin{eqnarray}}
\def\ea{\end{eqnarray}}
\def\etal{et al.\ \rm}

\title{Stellar growth by disk accretion: the effect of disk irradiation on the protostellar evolution.}

\author{Roman R. Rafikov\altaffilmark{1}}
\altaffiltext{1}{Department of Astrophysical Sciences, Princeton University, Ivy Lane, Princeton, NJ 08540, USA; rrr@astro.princeton.edu}


\begin{abstract}
Young stars are expected to gain most of their mass by accretion
from a disk that forms around them as a result of angular momentum 
conservation in the collapsing protostellar cloud. Accretion 
initially proceeds at high rates of $10^{-6}-10^{-5}$ M$_\odot$
yr$^{-1}$ resulting in strong irradiation of the stellar 
surface by the hot inner portion of the disk and leading to the 
suppression of the intrinsic stellar luminosity. Here we 
investigate how this luminosity suppression affects evolution 
of the protostellar properties. Using simple model
based on the energy balance of accreting star we demonstrate 
that disk irradiation causes only a slight increase of the 
protostellar radius, at the level of several per cent. Such a weak
effect is explained by a minor role played by the intrinsic stellar 
luminosity (at the time when it is significantly altered by 
irradiation) in the protostellar energy budget
compared to the stellar deuterium burning luminosity 
and the inflow of the gravitational potential energy 
brought in by the freshly accreted material. Our results justify 
the neglect of irradiation effects in previous 
studies of the protostellar growth via disk accretion. 
Evolution of some other actively  
accreting objects such as young brown dwarfs and planets should 
also be only weakly sensitive to the effects of disk irradiation.
\end{abstract}


\section{Introduction.}
\label{sect:intro}


It is currently established that circumstellar disks are quite 
ubiquitous around young stellar objects of all masses 
(Muzerolle \etal 2003; Cesaroni \etal 2007). They represent 
an important ingredient of the star formation since initially 
protostars must be growing predominantly by accretion through
the disk:
angular momentum conservation forces the infalling protostellar
cloud material to form a centrifugally supported disk which then 
accretes onto a star. It is quite likely that only a small 
fraction of the final stellar mass gets acquired by the direct 
infall onto the protostellar surface, so that almost all 
of the stellar mass gets processed through the disk.

This picture of star formation implies very high initial mass 
accretion rates in the disk, at the level of $10^{-6}-10^{-5}$ 
M$_\odot$ yr$^{-1}$ as the Solar-type stars are thought 
to gain most of their mass during the first several $10^5$ yr. 
The presence of such a high-$\dot M$ accretion flow just 
outside the protostar immediately raises an issue of its possible 
effect on the protostellar properties. 

There are several ways in which disk accretion affects 
the protostar. First, star gains mass from the disk which increases
its binding energy and tends to make the star more compact. Second, 
accreting gas brings in some amount of thermal energy with it which
contributes to the pressure support in the star. The exact amount 
of heat advected into the star with the accreted material is unknown but
it seems likely that because of the disk geometry the  
accreted gas would have enough time to radiate away most of its 
thermal energy and would join the convective interior of the star with 
temperature much smaller than the stellar virial temperature.

Third, intense energy dissipation taking place in the innermost parts
of the accretion disk leads to strong irradiation of the stellar
surface by the disk (Frank \& Shu 1986; Popham 1997). It has been 
recently realized (Rafikov 2007) that irradiation can act
to suppress the internal luminosity of the protostar similar to
the suppression of the cooling of hot Jupiters by the radiation of 
their parent stars (Guillot \etal 1996; Burrows \etal 2000; Baraffe 
\etal 2003; Chabrier \etal 2004). Disk accretion  
at rates $\dot M \sim 10^{-6}-10^{-5}$ M$_\odot$ yr$^{-1}$ 
can easily reduce internal stellar luminosity by a factor of several 
which may have important implications for the early stellar evolution.

The goal of the present paper is to assess these implications 
by following the evolution of accreting protostar properly 
taking into account effects of disk irradiation.


\section{Method of calculation.}
\label{sect:evolve}


We consider a protostar of mass $M$ and radius 
$R$ growing by accretion of gas at rate $\dot M$ from 
the circumstellar disk. In this work $\dot M$ is specified 
as an explicit function of time so that $M(t)$ is also 
known. We assume the disk to extend all the 
way to the stellar surface as even a $1$ kG magnetic field
(typical value measured in the mature T Tauri systems, see 
Bouvier \etal 2007) would not be able to truncate the disk 
accreting at a high rate of 
$\dot M \sim 10^{-6}-10^{-5}$ M$_\odot$ yr$^{-1}$. At the
stellar surface accreting gas passes through the boundary 
layer (Popham \etal 1993)
in which its speed is reduced from the Keplerian velocity
to the velocity of the stellar surface.

To evaluate the effect of disk irradiation on the protostellar 
evolution we use an approach based on the energy conservation 
which was developed in Hartmann \etal (1997). In this approach the 
convective part of the star comprising most of its mass is 
assumed to behave like a polytrope with index $n=3/2$, so that
inside the star pressure $P$ is related to the density $\rho$ via
$P\propto\rho^{5/3}$. This approximation works very well in
highly ionized, dense, and fully convective interiors of young
stars. The total energy of such a star (a sum of its thermal and 
gravitational energies\footnote{In this work we neglect stellar 
rotation.}) is $E_{tot}=-(3/7)GM^2/R$ 
(Kippenhahn \& Weigert 1994).   
Evolution of protostellar properties -- luminosity $L$, radius 
$R$ -- as a function of time [or, equivalently, stellar mass
$M(t)$] is then governed by the following equation:
\ba
\frac{d}{dt}\left(-\frac{3}{7}\frac{GM^2}{R}\right)=
-\frac{GM\dot M}{R}+\dot E_{th}+L_D-L.
\label{eq:energy_eq}
\ea
The l.h.s. of this equation represents the change in the total 
stellar energy, the first and second terms in the r.h.s. 
are the gravitational potential energy and the thermal energy 
brought in with the accreted material,
while $L_D(M, R)$ is a deuterium luminosity of a 
protostar.  Stellar luminosity $L$ is the luminosity
carried towards the photosphere by the convective motions in 
the stellar interior. It is different from the integrated 
emissivity of the stellar surface since the star also 
intercepts and reradiates a fraction of energy released in the 
accretion disk. 

Rate at which thermal energy gets accreted by the star is 
\ba
\dot E_{th}=\frac{\dot M}{\gamma-1}\frac{k_B T}{\mu}=\alpha 
\frac{GM \dot M}{R},
\ea
where $\gamma$ is the ratio of specific heats of accreted gas 
(which can be different from $\gamma=5/3$ characteristic for 
the stellar interior), $k_B$ is the Boltzmann constant, $\mu$
and $T$ are the mean molecular weight and the temperature of 
the accreted gas. Dimensionless parameter $\alpha$ can be 
written as
\ba
\alpha=\frac{1}{\gamma-1}\frac{T}{T_{vir}},~~~T_{vir}\equiv
\frac{\mu}{k_B}\frac{GM}{R},
\label{eq:alpha}
\ea
where $T_{vir}$ is the stellar virial temperature. Gas accreting from the
disk experiences strong dissipation in the boundary layer near 
the stellar surface. In this layer gas temperature can become
an appreciable fraction of $T_{vir}$. However, the cooling time
in the boundary layer and the outermost layers of the star is 
very short so that the accreted gas cools efficiently and should 
ultimately join stellar interior with temperature $T$
which is much lower than $T_{vir}$ [unless $\dot M$ is extremely 
high, in excess of $10^{-4}$ M$_\odot$ yr$^{-1}$, see Popham (1997)].
Thus, under the conditions considered in this work one
expect $\alpha\ll 1$ but the actual value of this parameter is 
rather poorly constrained (it depends on the gas thermodynamics, 
radiative transport,  and the 
viscosity prescription in the boundary layer which are poorly 
constrained). As here we are primarily interested 
in the effects of disk irradiation we set $\alpha=0$ for 
simplicity\footnote{Prialnik \& Livio (1985) and Hartmann 
\etal (1997) have previously investigated the effect of the 
variation of $\alpha$ on the protostellar evolution.}. In this
case equation (\ref{eq:energy_eq}) can be rewritten as
\ba
\frac{\dot R}{R}=\frac{7}{3}\frac{R}{GM^2}
\left(L_D-L\right)-\frac{1}{3}\frac{\dot M}{M}. 
\label{eq:master}
\ea
This is an evolution equation for $R$ and can be easily
integrated numerically once the dependencies of $L_D$ and 
$L$ on stellar parameters are known.

For $L_D$ we adopt the expression obtained in Stahler (1998)
by integrating the rate of energy release due to D burning 
within the $n=3/2$ polytrope:
\ba
L_D=f_D\rm{[D/H]}L_{D,0}\left(\frac{M}{M_\odot}\right)^{13.8}
\left(\frac{R}{R_\odot}\right)^{-14.8},
\label{eq:L_D}
\ea
where $L_{D,0}=1.92\times 10^{17}~L_\odot$ and $f_D$ is the 
fractional D abundance relative to the initial D number abundance  
[D/H] taken to be $2\times 10^{-5}$. Parameter $f_D$ is not 
constant in time -- it evolves since D burns in the stellar 
interior while the new D is being brought in with the accreting 
material 
(with the initial abundance [D/H]). As a result, one finds
(Stahler 1988; Hartmann \etal 1997)
\ba
\frac{d}{dt}(f_DM)=\dot M-\frac{L_D}{\beta_D},
\label{eq:f}
\ea 
where $\beta_D=9.2\times 10^{13}$ ergs g$^{-1}$ is the energy 
released by D fusion per gram of stellar material (assuming 
[D/H]$=2\times 10^{-5}$).

The most important aspect of this work which distinguishes it
from Hartmann \etal (1997) is the calculation of $L$.
In the approximation adopted by Hartmann \etal (1997) $L$
is a function of $R$ and $M$ only. In our case situation 
is different: irradiation of the stellar surface 
gives rise to an outer convectively stable layer below 
the stellar photosphere (Rafikov 2007), similar to the radiative layer
that form in the atmospheres of the close-in giant planets irradiated
by their parent stars (Guillot \etal 1996; Burrows \etal 2000). 
This external radiative zone suppresses
the local radiative flux coming from stellar interior
and this changes the integrated stellar luminosity, which becomes 
a function of irradiation intensity. As a result, in irradiated case 
$L$ depends not only on $R$ and $M$ but also on 
$\dot M$ (which determines the strength of the irradiation flux). 

Rafikov (2007) has demonstrated that for a given opacity 
behavior at the stellar surface (parametrized in his case 
to be a power-law 
function of gas pressure $P$ and temperature $T$, 
$\kappa\propto P^{\alpha}T^\beta$) the degree of luminosity 
suppression depends only on the so-called irradiation 
parameter 
\ba
L=L_0\chi(\Lambda),~~~\Lambda\equiv 4\pi
\frac{GM\dot M}{R L_0},
\label{eq:Lambda}
\ea 
which is (up to a constant factor) the ratio of the accretion 
luminosity of the disk $GM\dot M/R$ to the luminosity $L_0$ that 
a star would have had in the absence of irradiation. Suppression 
factor $\chi(\Lambda)\to 1$ as $\Lambda\ll 1$ while 
$\chi(\Lambda)\lesssim 1$ when $\Lambda\gg 1$. 
In a simple case considered by Rafikov (2007) the dependence 
of $\chi$ on the opacity behavior comes only through the 
parameter\footnote{For convection to set in at some depth 
below the outer radiative zone a condition $\xi>4$ must be 
fulfilled, see Rafikov (2007) for details.} 
\ba
\xi=\beta+(1+\alpha)/\nabla_{ad},
\label{eq:xi}
\ea
where $\nabla_{ad}$ is 
the  adiabatic temperature gradient near the stellar 
surface. 

In the strongly irradiated case the temperature of the stellar 
surface varies as a function of latitude: equatorial 
belt is strongly heated by the hot inner parts of the disk while 
the polar regions of the star are virtually unaffected by 
irradiation and preserve their temperature at the level of 
$T_0=(L_0/4\pi R^2\sigma)^{1/4}$. In this situation one 
may wonder whether it is reasonable to assume a fixed opacity law
(as was done in Rafikov 2007) for the calculation 
of luminosity suppression given that the behavior of $\kappa$ can be 
different between the polar and the equatorial regions of the star. 
However, is was shown in Rafikov (2007) that for the opacity 
behavior typical for stellar
photospheres in the temperature interval from $\sim 2.5\times 10^3$
K to $10^4$ K the cooling of irradiated stars occurs mainly 
through their polar caps\footnote{Situation is different in the 
case of giant planets irradiated by the circumplanetary disks, see 
Rafikov (2007) for details.} (even though the opacity scaling with 
temperature and pressure changes quite drastically within this 
temperature interval at around $5000$ K). As a result, no matter 
how hot the equatorial parts of the star become and how 
complicated the opacity behavior is in this portion of the 
stellar surface, the integrated stellar luminosity $L$ does  
not depend on these details very strongly but is  
rather determined by the properties of the polar regions 
of the star: the size of the cool polar caps in which the 
photospheric temperature is preserved at the level of $T_0$, 
and the opacity behavior in the adjacent parts of the 
stellar surface. This provides motivation for using the stellar
luminosity prescription represented by equation (\ref{eq:Lambda}).
In this paper we adopt $\kappa$ characteristic for the 
temperature interval $2.5\times 10^3$ K 
$\lesssim T\lesssim 5\times 10^3$ K (Bell \& Lin 1994):
\ba
\kappa\approx 6\times 10^{-14}P^{2/3}T^{7/3}.
\label{eq:opacity}
\ea
This scaling should be reasonable in the polar regions of 
irradiated stars where the photospheric temperature $T_0$ is 
not strongly affected by irradiation and is close to the 
photospheric temperature that an isolated non-accreting star 
would have possessed in the Hayashi phase. The opacity law 
(\ref{eq:opacity}) corresponds to $\xi=6.5$, assuming 
$\nabla_{ad}=2/5$ as appropriate for the fully ionized stellar
interior ($\gamma=5/3$). 
Having specified the opacity law we have thus fully determined 
the behavior of $\chi(\Lambda)$ (which we take from Rafikov [2007],
see the curve corresponding to $\xi=6.5$ in Fig. 4 of that paper) 
necessary to compute $L$.

One remaining ingredient of the calculation is the choice of 
$L_0(M,R)$ -- the luminosity of a non-irradiated star. 
Hartmann \etal (1997) have 
adopted the following fit to the stellar evolution tracks of
D'Antona \& Mazzitelli (1994):
\ba
L_0(M,R)=1 ~ L_\odot 
\left(\frac{M}{0.5~M_\odot}\right)^{0.9} 
\left(\frac{R}{2~R_\odot}\right)^{2.34}. 
\label{eq:L_0}
\ea
Evolution tracks in D'Antona \& Mazzitelli (1994) have been 
calculated using the equation of state from Magni \& Mazzitelli (1979) 
that has been superceded by the more refined treatments 
(Saumon \etal 1995).  Also, D'Antona \& Mazzitelli's treatment of 
convection is based on Canuto \& Mazzitelli (1991) which has 
previously raised some concerns (Demarque \etal 
1999; Nordlund \& Stein 1999). 
Despite these deficiencies, we have chosen to adopt the 
prescription (\ref{eq:L_0}) in our work because of its 
simplicity\footnote{We do not require perfect knowledge of $L_0$ 
since our primary goal is to evaluate the importance of disk 
irradiation.} and also to allow direct 
comparison with the results of Hartmann \etal (1997).

Equations (\ref{eq:master}), (\ref{eq:L_D}), (\ref{eq:f}), 
(\ref{eq:Lambda}), \& (\ref{eq:L_0}) 
supplemented with $dM/dt=\dot M(t)$ and 
the dependence $\chi(\Lambda)$ from Rafikov (2007)
fully determine the evolution of an accreting protostar irradiated
by its own disk. This system of equations 
is then evolved numerically assuming that the prescription for 
$\dot M(t)$ is given.

\begin{figure}
\plotone{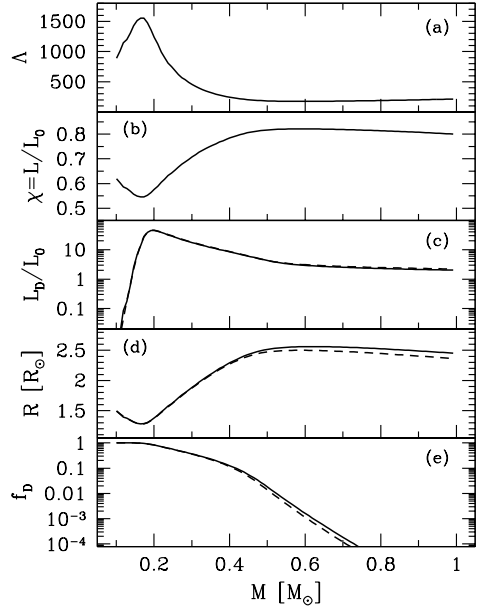}
\caption{
Evolution of the protostellar properties for $\dot M=4\times 10^{-6}$
M$_\odot$ yr$^{-1}$. Initially $M=0.1$ M$_\odot$ and 
$R=1.5$ R$_\odot$. We display the runs of irradiation parameter 
$\Lambda$ ({\it a}), suppression factor $\chi(\Lambda)$ ({\it b}), 
ratio $L_D/L_0$ of the D burning luminosity to the luminosity given 
by Eq. (\ref{eq:L_0}) ({\it c}), stellar radius $R$ ({\it d}), and 
the D abundance $f_D$ relative to its initial value ({\it e}), as 
functions of stellar mass $M$. In the last three panels we display 
the corresponding quantities both in irradiated ({\it solid}) and 
non-irradiated ({\it dashed}) cases.
\label{fig:f1}}
\end{figure}


\section{Results.}
\label{sect:res}


Here we present the results of our calculations. As initial 
conditions we choose $M=0.1$ M$_\odot$ and $f_D=1$. We vary
$R$ and the prescription for $\dot M(t)$ to see their 
effect on the evolution of stellar properties. 

In Figure \ref{fig:f1} we display protostellar evolution
for initial $R=1.5$ R$_\odot$ and a uniform $\dot M=4\times 10^{-6}$
M$_\odot$ yr$^{-1}$ with and without the effects of disk irradiation 
included. One can see that in both irradiated and non-irradiated 
cases evolution is pretty much the same: star initially contracts
until its central density and temperature become high enough for the 
deuterium to ignite. Right after that D burning strongly dominates 
over the stellar luminosity and D abundance $f_D$ starts going down. 
Resulting energy release in the stellar interior causes 
star to expand out to $2.5$ R$_\odot$ and D luminosity $L_D$ 
decreases appreciably (but still exceeds $L_0$ by a factor of 
several). These results are in full agreement with the calculations
of Hartmann \etal (1997).

In the top two panels of Figure \ref{fig:f1} we show quantities 
unique for the irradiated case: run of the irradiation parameter
$\Lambda$ and the suppression factor $\chi(\Lambda)$. As $R$
initially decreases $\Lambda$ goes up to $\approx 1.5\times 10^3$ since 
$\Lambda\propto \dot M M^{0.1}R^{-3.34}$ for $L_0$ given by 
equation (\ref{eq:L_0}). As a result, the internal stellar
luminosity is appreciably suppressed and $\chi$ reaches $\approx 0.55$ 
demonstrating the importance of disk irradiation in regulating $L$. 

\begin{figure}
\plotone{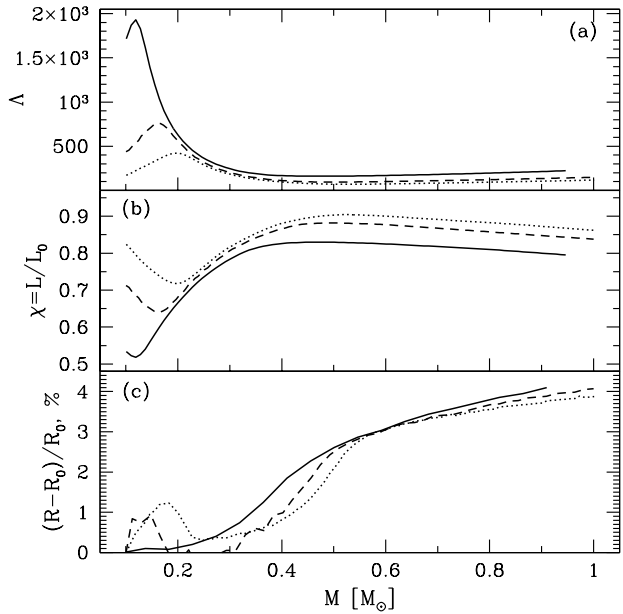}
\caption{
Relative difference between the stellar radii in the irradiated 
$R$ and non-irradiated $R_0$ cases ({\it panel c}) for a 
protostar accreting 
at a uniform rate $\dot M=2\times 10^{-6}$ M$_\odot$ yr$^{-1}$.
Different curves correspond to different initial radii: 
$1.0$ R$_\odot$ ({\it solid}), $1.5$ R$_\odot$ ({\it dashed}), 
and $2.0$ R$_\odot$ ({\it dotted}). Panels {\it a} and  {\it b}
display runs of irradiation parameter $\Lambda$ and suppression 
factor $\chi$ as a function of stellar mass.
\label{fig:f2}}
\end{figure}

At the same time, although the
effect of irradiation on $L$ is of order unity, Figure 
\ref{fig:f1} clearly demonstrates that irradiation affects other stellar
properties such as $R$ and $f_D$ only very weakly. As expected,
$R$ in irradiated case is larger than in the non-irradiated case 
(because $L$ suppression allows more heat to be 
retained inside the star) but only by several per cent.  

\begin{figure}
\plotone{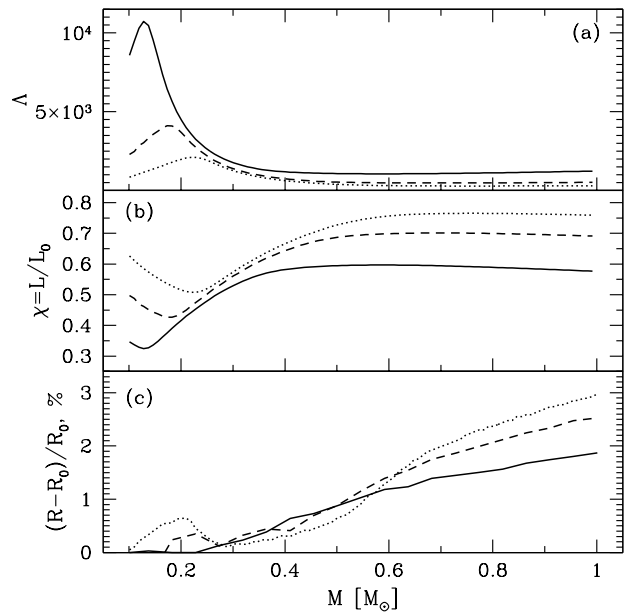}
\caption{
Same as Fig \ref{fig:f2} but for a uniform 
$\dot M=10^{-5}$ M$_\odot$ yr$^{-1}$.
\label{fig:f3}}
\end{figure}

To check that this result is not an artefact of our initial 
conditions and assumed $\dot M$ we have additionally  
calculated protostellar evolution for uniform $\dot M=2\times 10^{-6}$
M$_\odot$ yr$^{-1}$ and $\dot M=10^{-5}$ M$_\odot$ yr$^{-1}$ and different
initial $R$. Results are presented in Figures \ref{fig:f2} and 
\ref{fig:f3} in which we display the evolution of 
$\Delta R/R=(R-R_0)/R_0$, where $R$ and $R_0$ are the values of 
the protostellar radius with and without disk irradiation taken into 
account. It is quite clear from these plots that despite the rather 
severe luminosity suppression ($\chi$ reaching $0.3$ in the high 
$\dot M$ case for the initial $R=1$ R$_\odot$) the relative 
stellar radius increase due to irradiation is always rather small. 
Note that at $M\sim$ M$_\odot$ we find $\Delta R/R\approx 4\%$ when 
$\dot M=10^{-5}$ M$_\odot$ yr$^{-1}$ (Figure \ref{fig:f2}) 
which is larger than in the higher $\dot M$ case shown in Figure 
\ref{fig:f3} (when $\Delta R/R\approx 3\%$).

Very similar outcome has been found in the case of non-uniform 
$\dot M$: in Figure \ref{fig:f4} we plot $\Delta R/R$ for different
initial $R$ and $\dot M=a M$ with $a$ chosen in such a way that a 
protostar grows to $1$ M$_\odot$ in $10^5$ yr. This growth time is close to 
the time needed to reach $1$ M$_\odot$ in the case of constant 
$\dot M=10^{-5}$ 
M$_\odot$ yr$^{-1}$, see Figure \ref{fig:f3}.  For non-uniform $\dot M $
one again finds that irradiated protostar differs in radius from the 
non-irradiated star by only a few per cent
($\Delta R/R\approx 3\%$ at $M\sim$ M$_\odot$). It is also obvious from 
the comparison of Figures \ref{fig:f3}c and \ref{fig:f4}c that roughly 
the same growth time translates into very similar behavior of  
$\Delta R/R$ in the two cases independent of how $\dot M(t)$ evolves.
We thus conclude that irradiation by the circumstellar accretion disk 
has rather small effect on the protostellar evolution and that 
this outcome is rather generic. 

\section{Discussion.}
\label{sect:disc}


Weak sensitivity of $R$ to disk irradiation 
given the strong effect that irradiation has on the internal 
luminosity $L$ of a protostar may seem surprising. However, one 
should bear in mind that besides $L$ there are other contributors 
to the stellar energy budget, namely the D burning luminosity 
$L_D$ and the gravitational potential energy gained with 
the accreted material. It turns out that these contributions 
dominate the energy budget over $L$. 

To see this we rewrite equation 
(\ref{eq:master}) using definition 
(\ref{eq:Lambda}) in the following form:
\ba
\frac{\dot R}{R}=\frac{28\pi}{3}\Lambda^{-1}
\frac{\dot M}{M}\left[\frac{L_D}{L_0}-\chi(\Lambda)-
\frac{\Lambda}{28\pi}\right]. 
\label{eq:rewr_master}
\ea
In the right-hand side of this equation the first term in 
brackets describes the relative role of $D$ burning in the 
total energy budget, second term represents intrinsic stellar 
luminosity, while the third term is due to the inflow of 
the gravitational potential energy $GM\dot M/R$ --
the ratio of this energy inflow to $L_0$ differs from $\Lambda$ 
only by a constant factor.

\begin{figure}
\plotone{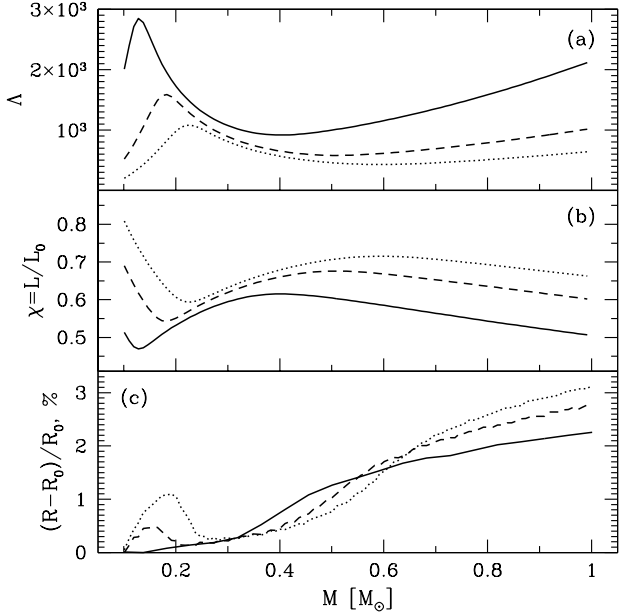}
\caption{
Same as Fig \ref{fig:f2} but for $\dot M\propto M$ and the growth 
time $10^5$ yr to reach $M=1$ M$_\odot$.
\label{fig:f4}}
\end{figure}

Suppression of $L$ is largest when $\Lambda$ is highest which
is easy to see by inspecting Figures \ref{fig:f1}-\ref{fig:f4}. But
this automatically means that the maximum deviation of the 
suppression factor $\chi$ from unity occurs precisely when the inflow of 
the gravitational potential energy far exceeds the stellar luminosity. 
Apparently, under these circumstances $L$ is a subdominant 
contribution to the stellar energy budget and thus even a 
significant reduction of $L$ compared to $L_0$ is going to be 
negligible compared to the gravitational energy influx. 

Moreover, Figure \ref{fig:f1} demonstrates that 
$\Lambda$ reaches its maximum when $R$ is at its 
minimum while $f_D$ is still very close to unity. At this point
vigorous D burning commences inside the star giving rise to
very high $L_D/L_0$. As a result, at the evolutionary stage 
when $\chi$ is minimal $L$ is subdominant in comparison to 
not only $GM\dot M/R$ but also $L_D$. This additionally downplays 
the role of the luminosity suppression by irradiation in the 
early protostellar evolution. 

This line of reasoning also explains why at $M\sim 1$ M$_\odot$ 
we have found $\Delta R/R$ to be larger for lower 
$\dot M$ (see \S \ref{sect:res}). First, smaller $\dot M$ 
means lower $\Lambda$ so that the ratio of $L$ to the 
gravitational energy inflow rate $GM\dot M/R$ in the low $\dot M$ 
case is larger than in the high $\dot M$ case. Also,
at $M\sim$ M$_\odot$ one generally finds $f_D\ll 1$ (see Figure
\ref{fig:f1}e) so that $L_D$ is mainly due to the burning of the 
freshly accreted D (rather than the D that remained in the protostar 
from previous accretion). Since in the low $\dot M$ case less fresh 
$D$ is supplied to the protostar $L_D$ must also be lower than in the 
high $\dot M$ case. As a result, in the lower $\dot M$ case 
$L$ plays a more significant role compared to $L_D$ (in which case 
$\Delta R/R$ should be more sensitive to changes in $L$ caused by 
irradiation) than in the high $\dot M$ case. 

\begin{figure}
\plotone{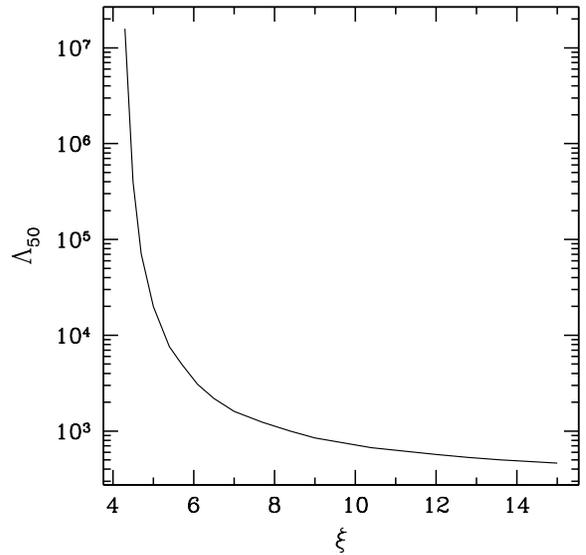}
\caption{
Plot of $\Lambda_{50}$ (the value of $\Lambda$ at which 
the suppression factor $\chi=0.5$) as a function of parameter 
$\xi$ characterizing the opacity law in the outer layers of
irradiated protostar. Note that $\Lambda_{50}$ never decreases 
below several hundred.
\label{fig:f5}}
\end{figure}

This conclusion immediately raises the following question: since 
$\Delta R/R$ increases as $\dot M$ decreases  would
one find $\Delta R/R\sim 1$ at low enough $\dot M$? The answer 
is no, and it has to do with the fact that 
$\chi$ appreciably differs from unity (obviously, a necessary 
condition for getting $\Delta R/R\sim 1$) only at rather high 
$\Lambda$. This is a generic feature of disk irradiation which 
is illustrated in Figure \ref{fig:f5} where we display 
$\Lambda_{50}$ -- the value of $\Lambda$ at which 
$\chi(\Lambda_{50})=0.5$ -- as a function of assumed opacity law
represented by the parameter $\xi$, see equation (\ref{eq:xi}).
One can see that $\Lambda_{50}\gtrsim 10^2$ for all $\xi>4$, meaning 
that significant luminosity suppression requires rather high $\dot M$.
This inefficiency of irradiation in suppressing $L$ 
is caused by the specific geometry of disk irradiation in which
the irradiation flux is a very sensitive function 
($\propto \theta^5$) of the latitude at the stellar surface 
$\theta$, see Rafikov (2007). Because of that stellar polar caps 
can stay cool even at rather high $\dot M$ allowing unsuppressed 
flux to be emitted over a significant portion of the stellar surface.

In the case of $\xi=6.5$ as appropriate for cool, low-mass 
protostars one finds that $\Lambda_{50}=2.2\times 10^3$ which 
according to equation (\ref{eq:Lambda}) immediately implies 
that $GM\dot M/R\approx 175 L_0$ when $\chi=0.5$. Clearly, in this
case stellar luminosity should have small effect on the protostellar 
evolution. If $\dot M$ is 
reduced\footnote{At small enough $\dot M$ ($\dot M\lesssim 10^{-6}$ 
M$_\odot$ yr$^{-1}$) protostellar magnetic field is likely to disrupt
accretion flow (K\"onigl 1991) thus invalidating our assumption of 
the direct disk accretion onto the star.} 
so that $GM\dot M/R\sim L_0$ (and $\Lambda\sim 1$) stellar luminosity 
would be playing a significant role in the stellar energy budget, 
however $\chi$ would be very close to unity (see Rafikov 2007) and 
the $L$ suppression by irradiation would be negligible. Thus, 
under no circumstances should one expect $\Delta R/R$ larger than 
several per cent, meaning that quite generally the irradiation by 
accretion disk is unlikely to play a significant role in the 
evolution of the protostellar properties.

Looking at this conclusion under slightly different angle, 
our results also imply that when considering evolution 
of the protostars accreting at $\dot M\sim 10^{-6}-10^{-5}$ 
M$_\odot$ yr$^{-1}$ one may completely neglect 
$L$ in the stellar energy budget and still get rather
decent description of the protostellar evolution. 

Our results are obtained assuming a specific value of $\xi$ 
typical for the low-mass stars. However, one expects our major 
conclusions to remain valid also in the case of 
other accreting objects such as young brown dwarfs and giant 
planets. Although these objects are likely to be characterized by  
values of $\xi$ different from $6.5$, our Figure \ref{fig:f5} 
clearly demonstrates that the efficient suppression of stellar 
luminosity requires very large $\Lambda$ for {\it any} value 
of $\xi$. As a result, we generically expect $L$
to be a subdominant energy contributor for any actively 
accreting object meaning that its radius (and 
other properties) should be different from the radius of its 
non-irradiated counterpart by at most a few 
per cent.

Previous evolutionary studies of the protostellar assembly by 
accretion (Mercer-Smith \etal 1984; Palla \& Stahler 1992; 
Siess \& Forestini 1996; Hartmann \etal 1997; Siess \etal 1997, 
1999) have always neglected the effect of disk 
irradiation on stellar properties. Results of this work demonstrate
that this simplification should not have led to major deviations from 
reality since disk irradiation affects stellar properties only at 
the few per cent level. Protostellar evolution may more sensitively 
depend\footnote{As long as $\alpha$ is not very 
small. Whether this is true is debatable 
(Popham 1997), and we assume $\alpha=0$ in this work.} 
(Prialnik \& Livio 1985; 
Hartmann \etal 1997) on the exact amount of thermal energy 
that gets accreted with the disk material which is parametrized by the
parameter $\alpha$, see equation (\ref{eq:alpha}). Disk 
irradiation is likely to play a very important 
role is setting the value of $\alpha$ since the 
conditions in the external radiative zone near the stellar 
surface regulate the thermodynamic properties of the
gas at the convective-radiative boundary of the protostar, which
is likely to be crucial in determining $\alpha$.


\section{Summary.}
\label{sect:concl}


We considered evolution of the properties of a growing 
protostar accreting gas from the circumstellar disk
with the goal of assessing the impact of irradiation 
by the inner regions of the disk. We find that disk
irradiation plays minor role in the radius evolution 
of a protostar so that the radius of an irradiated star 
is larger by only a few per cent compared to the non-irradiated 
case. This result is largely independent of
the specific behavior of the protostellar mass accretion 
rate as long as $\dot M$ is high enough, at the level of 
$10^{-6}-10^{-5}$ M$_\odot$ yr$^{-1}$. The weak sensitivity 
of the stellar properties to the disk irradiation is explained 
by the minor role played by the stellar luminosity $L$ in the energy
budget of the star at the time when $L$ is significantly altered 
by irradiation. At these stages of protostellar evolution the 
inflow of the negative gravitational potential energy brought 
in by the accreting material and the deuterium burning luminosity 
are the much more significant contributors to the stellar energy 
budget. Our results imply that the previous studies of the protostellar 
accretion which neglected disk irradiation may have underpredicted 
stellar radius only at the level of several per cent. Our major 
conclusions should be directly applicable to the evolution of 
other actively accreting objects such young brown dwarfs and planets.

\acknowledgements 



\end{document}